\documentstyle[aps,prb,preprint]{revtex}
\begin{document}

\title{Electrostatic potentials and free energies of solvation of
  polar and charged molecules}

\author{Gerhard Hummer,$^{*}$ Lawrence R. Pratt, Angel E.
  Garc\'{\i}a, Bruce J. Berne$^1$ and Steven W. Rick$^2$}

\address{Theoretical Division, MS K710, Los Alamos National
  Laboratory, Los~Alamos, New Mexico 87545\\
  Phone: (505) 665-1923; Fax: (505) 665-3493; e-mail:
  hummer@lanl.gov\\
  $^1$ Department of Chemistry and Center for Biomolecular
  Simulation, Columbia University, New York, New York 10027\\
  $^2$Frederick Cancer Research and Development Center, National
  Cancer Institute, Frederick, Maryland 21702}

\date{in press: {\em The Journal of Physical Chemistry};
  LA-UR 96-4479}

\maketitle

\begin{abstract}Theories of solvation free energies often involve
  electrostatic potentials at the position of a solute charge.
  Simulation calculations that apply cutoffs and periodic boundary
  conditions based on molecular centers result in center-dependent
  contributions to electrostatic energies due to a systematic sorting
  of charges in radial shells. This sorting of charges induces a
  surface-charge density at the cutoff sphere or simulation-box
  boundary that depends on the choice of molecular centers.  We
  identify a simple solution that gives correct, center-independent
  results, namely the radial integration of charge densities.  Our
  conclusions are illustrated for a Lennard-Jones solute in water.
  The present results can affect the parameterization of force
  fields.
\end{abstract}

\clearpage

Accurate simulation calculations of free energies of solvation require
a careful treatment of long-range electrostatic interactions.  Recent
computational and theoretical work on single-ion free
energies\cite{Guggenheim:67} has converged upon a common set of ideas
that are, however, discussed in slightly different ways, {\em i.e.},
Gaussian fluctuations of electrostatic potentials,\cite{Levy:91}
second-order perturbation theory,\cite{Pratt:94:a} or linear-response
theory.\cite{Jayaram:89,Aqvist:96} These approaches require the
calculation of electrostatic potentials at atom positions on a solute
molecule at fractional charge states ({\em e.g.}, uncharged or fully
charged).  However, a lack of consensus on how electrostatic
potentials should be evaluated means that calculated partial
contributions to single-ion free energies are often not fully
comparable.  Differences arise because of a common practice of
evaluating electrostatic interactions considering whole molecules.
This can lead to spurious dependences on the choice of the center of a
molecule.  Similar issues arose in calculations of the electrostatic
potential difference of the water-vapor interface: seemingly identical
calculations of electrostatic potentials can produce different final
results.\cite{Wilson:89}

Discrepancies in calculated electrostatic potentials were noted
recently by {\AA}qvist and Hansson.\cite{Aqvist:96} The present paper
resolves the difficulties noted there.  We will focus on the
calculation of electrostatic potentials at the position of a solute
molecule in a polar fluid, discussing the effects of different methods
of summing charge interactions.  This leads us to a simple, center
independent, and feasible recipe used to analyze electrostatic
potentials, both in finite and infinite systems, namely spherical
integration of charge densities.  To illustrate our general results,
we will show data for Lennard-Jones (LJ) solutes in water.

Two different center dependences will be considered (see
Figure~\ref{fig:scheme}).  The first is associated with the center of
the solvent molecule denoted by M used to bin electrostatic
interactions between solvent molecules and the solute molecule.  The
second center dependence to consider is the dependence on the solvent
center P that might be used in implementing minimum-image periodic
boundary conditions (PBC's) by translating a whole solvent molecule
into the primary simulation box.  These two centers M and P would
often coincide but they need not.  The effects considered are
distinct.

For molecule-based summation, the electrostatic potential at the
center of a spherical LJ solute molecule depends strongly on the
choice of the center M of a water molecule that defines into which
shell it belongs.  Shown in Figure~\ref{fig:figpot} are curves
$\phi_M(r)$ of potential contributions of water molecules with their
center M within a radius $r$ of the solute molecule,
\begin{eqnarray}
  \label{eq:phimol}
  \phi_M(r) & = & \int_0^r dr\left\langle \sum_{i=1}^N
    \delta(r-r_{i,M}) \sum_{\alpha=1}^3 \frac{q_\alpha}{r_{i,\alpha}}
  \right\rangle~,
\end{eqnarray}
where the $\alpha$ sum extends over the water oxygen atom O and
hydrogen atoms H1 and H2.  $\langle\ldots\rangle$ denotes a canonical
ensemble average over a system of $N$ SPC water
molecules\cite{Berendsen:81} with oxygen and hydrogen positions ${\bf
  r}_{i,O}$, ${\bf r}_{i,H1}$ and ${\bf r}_{i,H2}$, respectively
($r_{i,O}=|{\bf r}_{i,O}|$, {\em etc.}); and one uncharged LJ solute
atom at position ${\bf r}_S=0$ with SPC-water LJ parameters.
$\delta(r)$ is the Dirac delta function.  $q_O$ and $q_H$ are the
charges on the oxygen and hydrogen sites of SPC water ($-0.82e$ and
$0.41e$, respectively).  ${\bf r}_{i,M}$ is the center of water
molecule $i$, defined as ${\bf r}_{i,M} = w {\bf r}_{i,O} + ( 1 - w )
( {\bf r}_{i,H1} + {\bf r}_{i,H2} )/2$.  The atom positions ${\bf
  r}_{i,O}$, ${\bf r}_{i,H1}$ and ${\bf r}_{i,H2}$ are shifted
molecule-based under PBC's.  (That is, the center P=M is mapped into
the simulation box, leaving the molecule intact so that individual
atoms can actually be outside the simulation box.)  For weights $w=1$
and 0, the center position ${\bf r}_{i,M}$ coincides with the oxygen
position and the hydrogen bisector, respectively.

The molecule-based potential defined in eq~\ref{eq:phimol} contrasts
with the charge-based potential $\phi_q(r)$:
\begin{mathletters}
  \label{eq:cha}
  \begin{eqnarray}
    \label{eq:phicha}
    \phi_q(r) & = & 4\pi \int_0^r r^2 dr\; \rho_q(r) / r~,\\
    \label{eq:rhoq}
    \rho_q(r) & = & \left\langle \sum_{i=1}^N \sum_{\alpha=1}^3
      q_\alpha (4\pi r^2)^{-1} \delta(r-r_{i,\alpha}^s) \right\rangle~.
  \end{eqnarray}
\end{mathletters}
$\rho_q({\bf r})$ is the radially averaged charge density.  In
eq~\ref{eq:rhoq}, PBC's for the positions of charges ${\bf
  r}_{i,O}^s$, ${\bf r}_{i,H1}^s$ and ${\bf r}_{i,H2}^s$ are applied
on the basis of atoms rather than molecules.

Each of the $\phi_M(r)$ curves in Figure~\ref{fig:figpot} for
different centers M reaches a plateau value after 0.6 to 0.8~nm
distance from the solute.  However, the plateau values differ not only
in magnitude but also in sign for different choices of M, whereas
identical choices of M give agreement between simulations under PBC's
and using clusters with 256 and 1024 water molecules.  The differences
are caused by the M-dependent sorting of molecules, even for identical
configurations (positions and orientations) of the solvent molecules.
If the center M is close to the oxygen atom, the first layer of
molecules considered in the integration in eq~\ref{eq:phimol}
predominantly includes water molecules with the oxygen atoms facing
the solute.  Correspondingly, $\phi_M(r)$ starts out negative as
negative contributions of the oxygen atoms dominate.  On the other
hand, if the center M is close to the hydrogen atoms, the first layer
of molecules considered in the integration will predominantly have the
hydrogen atoms facing the solute (see Figure~\ref{fig:scheme},
middle).  As a consequence, $\phi_M(r)$ starts out positive and also
reaches a positive plateau value.  Results for centers M between the
oxygen and the hydrogen bisector fall between the two curves.

For a finite sample, the different curves all converge to the same
value when all contributions have been summed up
(Figure~\ref{fig:figpot}, middle and bottom).  Convergence is
therefore reached only after crossing the interface to the exterior,
so that surface-potential contributions are included.  For the cluster
simulations of Figure~\ref{fig:figpot} (middle and bottom), the
potential crosses a liquid-vacuum interface.

Similar problems arise with molecule-based cutoffs (or residue-based
cutoffs for macromolecules).  For instance, if the distance to the
oxygen atom of a water molecule is used to determine whether a
particle interacts with that water molecule, a characteristic
surface-charge density is induced at the cutoff sphere.  The oxygen
density seen by the solute is essentially a step function.  The
hydrogen density is reduced just inside the cutoff and nonzero just
beyond the cutoff, resulting in a net negative charge density just
inside the cutoff sphere and a net positive charge density just
outside.  This effective surface-dipole density strongly affects the
potential at the site of the particle.  That effect is {\em
  independent} of the cutoff length, as the surface area and
charge-dipole interaction vary with the square and the inverse square
of the cutoff length, respectively.

When whole molecules are shifted under PBC's this leads to another
level of ill-definition of electrostatic potentials.  Shifting
molecules as a whole means that PBC's are applied based on a center P
with coordinates ${\bf r}_{i,P} = u {\bf r}_{i,O} + ( 1 - u ) ( {\bf
  r}_{i,H1} + {\bf r}_{i,H2} )/2$ with weight $u$.  If that center P
coincides with the center M of the $\phi_M$ integration, then the
plateau in $\phi_M(r)$ reached after the first layer of water
molecules remains essentially unchanged when reaching the box
boundary.  However, if P and M do not coincide, $\phi_M(r)$ crosses
over from the M curve to the P curve when the box boundary is reached.
This can be seen in Figure~\ref{fig:figpot} (top) where the
$\phi_M(r)$ curve for M equal to O ($w=1, u=0$) crosses over to the
hydrogen-bisector curve ($w=0, u=0$) when the hydrogen-bisector is
used for PBC's (P=HH).  Clearly, this is an unphysical behavior
associated with summing electrostatic interactions and applying PBC's
on the basis of molecules.

How can we eliminate these difficulties of calculating electrostatic
potentials in computer simulations?  The unphysical false plateaus
observed for $\phi_M(r)$ in Figure~\ref{fig:figpot} stem from
associating partial charges with molecular centers.  By choosing a
center M, the water molecules were systematically sorted for analysis.
For a {\em finite} system, integration to infinity is required to get
the correct result.  And that result will then contain troublesome and
undesired surface-potential contributions.  Under PBC's, that
integration cannot be performed easily, as is manifest from the
dependence of the limiting value of the potential on the choice of the
molecular center P upon which PBC's are applied.

However, if we alternatively integrate over {\em charge densities}
$\rho_q({\bf r})$ rather than sum over {\em molecules}, we will obtain
a well-defined result for the potential that coincides with taking the
limit of an infinite system before extending the integral to infinity.
The charge-based potential $\phi_q(r)$ is defined in
eq~\ref{eq:phicha}.  For a finite system, eqs~\ref{eq:phimol} and
\ref{eq:cha} will give identical results if the integration volume
covers the whole system (extending beyond the interface to the
container, vacuum {\em etc.}).  However, unlike eq~\ref{eq:phimol} the
potential $\phi_q(r)$ defined in eq~\ref{eq:phicha} will reach a
plateau beyond the correlation length of the charge correlation
$\rho_q(r)$ independent of an arbitrary choice of the center M of a
molecule.  (As shown in Figure~\ref{fig:figpot}, that plateau is
reached within about 1~nm from the neutral LJ solute.  Larger
correlation lengths were observed for a charged
solute.\cite{Hummer:96:a})

These issues would be largely irrelevant with conventional Ewald
treatment of electrostatic potentials, where the simulation box is
replicated periodically in space.  However, center dependences can
arise with modifications of the standard Ewald approach.  The
electrostatic potentials of periodic images can be summed up using the
Ewald potential $\varphi_E({\bf r})$.\cite{Hummer:96:a,Figueirido:95}
$\varphi_E({\bf r})$ is the periodic solution of Poisson's equation
$\Delta \varphi_E({\bf r}) = -4\pi [ \delta({\bf r})-1/V ] $ for a
unit point charge and a homogeneous background in the unit cell $V$.
The equivalents of the electrostatic potentials $\phi_M(r)$ and
$\phi_q(r)$ defined in eqs~\ref{eq:phimol} and \ref{eq:phicha} for
periodic systems are then
\begin{mathletters}
  \begin{eqnarray}
    \label{eq:phiEmol}
    \phi_M^E(r) & = & \int_0^r dr\left\langle \sum_{i=1}^N
      \delta(r-r_{i,M}) \sum_{\alpha=1}^3
        q_\alpha \varphi_E({\bf r}_{i,\alpha}) \right\rangle,\\
    \label{eq:phiEcha}
    \phi_q^E(r) & = & \int_0^r dr\left\langle \sum_{i=1}^N
      \sum_{\alpha=1}^3 \delta(r-r_{i,\alpha}^s)
      q_\alpha \varphi_E({\bf r}_{i,\alpha}^s) \right\rangle.
  \end{eqnarray}
\end{mathletters}
Again, minimum-image PBC's for ${\bf r}_{i,\alpha}$ and ${\bf
  r}_{i,\alpha}^s$ are applied on the basis of molecular centers P and
individual atoms, respectively.  Figure~\ref{fig:figEw} shows that the
charge-based Ewald potential and $1/r$ curves $\phi_q^E(r)$ and
$\phi_q(r)$ converge but that the molecule-based curve $\phi_M^E(r)$
for periodic systems also converges to $\phi_q^E(r)$ rather than
$\phi_M(r)$.  This is expected because the Ewald potential is fully
periodic.

Physical modification of the Ewald potential sacrifice this
periodicity.  The Ewald potential is the limit of performing the
lattice sum with the growing lattice embedded in a sphere cut out of a
medium with infinite dielectric constant $\epsilon'=\infty$ (tin-foil
boundary conditions).  Total potential energies without the effect of
that dielectric background $\epsilon'=\infty$ require subtraction of a
term proportional to the square of the net dipole moment {\bf M} of
the simulation box.\cite{deLeeuw:80:a} Expressed as an effective
potential, we can subtract a term $2\pi r^2/3V$ from $\varphi_E({\bf
  r})$: $\varphi_{E,\epsilon'=1}({\bf r}) = \varphi_{E}({\bf r})-2\pi
r^2/3V$.  This destroys the periodicity.  Use of the modified
potential $\varphi_{E,\epsilon'=1}({\bf r})$ in eq~\ref{eq:phiEmol}
forces $\phi_M^E(r)$ to converge to $\phi_M(r)$, as shown in
Figure~\ref{fig:figEw}.  However, the result for the potential
$\phi_M^E(r)$ at the solute site then again depends on the particular
choice of the molecular center P upon which PBC's are applied.
Clearly, to reproduce the non-physical effects of integrating the
potential using $1/r$ with molecule-based sorting requires subtraction
of a non-periodic term from the Ewald potential and application of the
potential outside the ``universe,'' {\em i.e.}, the simulation box.

It must be noted that subtracting the $r^2$ term from the Ewald
potential has little effect if the integration is based on charges
(Figure~\ref{fig:figEw}).  However, applying the $r^2$ modification by
molecules rather than atoms leads to large differences in the
potential. The charge-based potential (eq~\ref{eq:phiEmol}) with P=O
molecular-based shifting of the $r^2$ term has a value different in
sign from the charge-based curves shown in
Figure~\ref{fig:figEw}.\cite{Rick:94} We also emphasize that changing
the dielectric background to a finite value $\epsilon'<\infty$ in the
Ewald sum should not affect the charging of an ion at the center of
the box.  The dipolar field induced by a background $\epsilon'$ beyond
a spherical cavity around ${\bf r}=0$ is proportional to ${\bf
  r}\cdot{\bf M}$ which is zero at the position ${\bf r}=0$ of the
uncharged particle.  When a point multipole is charged from zero, that
contribution is also zero because of averaging over all orientations.

The results of this paper explain the differences in the sign of the
electrostatic potential at the position of an uncharged LJ particle in
water between {\AA}qvist and Hansson\cite{Aqvist:96} (M=O based
sorting, $1/r$: negative potential), Rick and Berne\cite{Rick:94}
(charge-based sorting; Ewald and $r^2$ modification with P=O based
shifting: negative potential) and Pratt {\em at al.}\cite{Pratt:94:a}
as well as Hummer {\em et al.}\cite{Hummer:96:a} (charge based
sorting; Ewald, $1/r$ and a generalized reaction-field interaction:
positive potential).  The best current value for that potential is
{\em positive.}  In that context a re-examination of several results
regarding free energies of charged species might be worthwhile.  For
instance, free energies of anions were found to be less negative in
Ref.~\onlinecite{Straatsma:88} than in Ref.~\onlinecite{Hummer:96:a}
but more negative for cations.  That can be explained if
molecule-based summation has been used in
Ref.~\onlinecite{Straatsma:88} using a center M at or close to the
oxygen atom of water.  The present results also affect the
parameterization of force fields involving charged species.  Finally,
we emphasize that the errors induced by molecule-based summation are
{\em independent} of the cutoff length for sufficiently large cutoffs.
If the induced surface-charge distribution were symmetrically
distributed on a spherical shell then it follows from Gauss's law that
the correction to the induced electrostatic potential inside the
spherical shell would be a constant.  In that case, the contributions
of M-dependent sorting would cancel each other for an overall neutral,
polar solute but not for a solute with a net charge.

Our results suggest that these issues are primarily matters of
analysis of configurational simulation data.  A variety of methods may
be used to obtain the configurational data.  The center dependences
considered here are introduced by the analysis of electrostatic
potentials and are often larger than the secondary differences in the
configurational data due to variations in their production.

The following general recipe for electrostatic-potential calculations
emerges: (1)~Electrostatic interactions should be integrated based on
{\em charge densities} rather than individual molecules to give
correct results for atoms and molecules carrying point charges or
spatially extended charge distributions.  For molecule-based
summation, the calculated potentials $\phi(r)$ level out nicely but
the plateau values depend on the arbitrary choice of molecular centers.
(2)~In simulations using PBC's, all charges should be mapped into the
simulation box.  Molecule-based PBC's result in center-dependent
surface-charge densities. (3)~Under PBC's, Ewald summation provides an
accurate way of summing up all interactions, minimizing finite-size
effects.

\paragraph*{Note added in proof.}
Ashbaugh and Wood\cite{Ashbaugh:96} come to similar conclusions
regarding molecule-center dependences of electrostatic potentials in
their comparison of Ewald summation\cite{Hummer:96:a} and cutoff
calculations.\cite{Wood:95} In particular, these authors also find the
potential to be positive for a neutral LJ solute in water.

  \begin{figure}[htbp]
    \caption{M- and P-center sorting of molecular partial
      charges. Left: Different M-centers considered for the water
      molecule.  M and M' coincide with the hydrogen bisector and the
      oxygen position.  Middle: Charges of the upper molecule are
      counted in the shaded spherical shell (bin) but not the charges
      of the lower molecule.  The lower molecule with an
      outward-pointing dipole moment is placed in a more distant bin.
      Right: P-center sorting, where P coincides with the oxygen
      position.  The bottom image of the molecule is considered in the
      electrostatic potential calculations.  For the particular choice
      of P=O and isotropic molecular orientations, the charge density
      is depleted just inside the simulation cell around the solute
      (outlined as square and circle, respectively) and enriched just
      outside.}
    \label{fig:scheme}
  \end{figure}

  \begin{figure}[htbp]
    \caption{Integrated electrostatic potentials at the position of an
      uncharged LJ solute in SPC water using $1/r$ interactions.
      Results are shown for different ways of sorting the charges and
      applying PBC's (atom or molecule based).  The top panel shows
      the results of averaging over 140\,000 Monte-Carlo passes of a
      system with 255 SPC water molecules and one LJ solute with
      SPC-water LJ parameters (using Ewald summation; see
      Ref.~\protect\onlinecite{Hummer:95:e} for simulation details).
      M and P denote the centers of sorting and applying PBC's,
      respectively, where O is the oxygen and HH the hydrogen-bisector
      position.  The middle and bottom panel show the results of
      averaging over 100\,000 and 300\,000 Monte-Carlo passes of
      clusters of 256 and 1024 SPC water molecules, respectively, and
      one LJ particle at the center, again with SPC-water LJ
      parameters.  In the cluster simulations, electrostatic
      interactions were calculated using $1/r$ Coulomb interactions
      without cutoff. The asymptotic value of charge-based integration
      using the Ewald potential is shown for reference.}
    \label{fig:figpot}
  \end{figure}

  \begin{figure}[htbp]
    \caption{Integrated electrostatic potential at the position of an
      uncharged LJ solute in SPC water using the Ewald potential
      $\varphi_E({\bf r})$ instead of $1/r$.  Results are shown for
      charge and molecule-based integration with and without the $r^2$
      modification added to $\varphi_E({\bf r})$.  See
      Figure~\protect\ref{fig:figpot} for further details.}
    \label{fig:figEw}
  \end{figure}


\begin{thebibliography}{10}

\bibitem{Guggenheim:67}
Guggenheim, E.~A.  {\em Thermodynamics. An Advanced Treatment for Chemists and
  Physicists};  John Wiley {\&} Sons: New York, 1967; Chap.~8.02 and 9.03.

\bibitem{Levy:91}
Levy, R.~M.; Belhadj, M.; Kitchen, D.~B. {\em J. Chem. Phys.} {\bf 1991}, {\em
  95},  3627.

\bibitem{Pratt:94:a}
Pratt, L.~R.; Hummer, G.; Garc\'{\i}a, A.~E. {\em Biophys. Chem.} {\bf 1994},
  {\em 51},  147.

\bibitem{Jayaram:89}
Jayaram, B.; Fine, R.; Sharp, K.; Honig, B. {\em J. Phys. Chem.} {\bf 1989},
  {\em 93},  4320.

\bibitem{Aqvist:96}
{\AA}qvist, J.; Hansson, T. {\em J. Phys. Chem.} {\bf 1996}, {\em 100},  9512.

\bibitem{Wilson:89}
Wilson, M.~A.; Pohorille, A.; Pratt, L.~R. {\em J. Chem. Phys.} {\bf 1989},
  {\em 90},  5211.
  
\bibitem{Berendsen:81} Berendsen, H. J.~C.; Postma, J. P.~M.; {van
    Gunsteren}, W.~F.; Hermans, J. In {\em Intermolecular Forces:
    Proceedings of the 14th Jerusalem Symposium on Quantum Chemistry
    and Biochemistry}; Pullman, B., Ed.; Reidel: Dordrecht, Holland,
  1981; pp\ 331--342.

\bibitem{Hummer:95:e}
Hummer, G.; Pratt, L.~R.; {Garc\'{\i}a}, A.~E. {\em J. Phys. Chem.} {\bf 1995},
  {\em 99},  14188.

\bibitem{Hummer:96:a}
Hummer, G.; Pratt, L.~R.; {Garc\'{\i}a}, A.~E. {\em J. Phys. Chem.} {\bf 1996},
  {\em 100},  1206.

\bibitem{Figueirido:95}
Figueirido, F.; {Del Buono}, G.~S.; Levy, R.~M. {\em J. Chem. Phys.} {\bf
  1995}, {\em 103},  6133.

\bibitem{deLeeuw:80:a}
de~Leeuw, S.~W.; Perram, J.~W.; Smith, E.~R. {\em Proc. R. Soc. London A} {\bf
  1980}, {\em 373},  27.

\bibitem{Rick:94}
Rick, S.~W.; Berne, B.~J. {\em J. Am. Chem. Soc.} {\bf 1994}, {\em 116},  3949.

\bibitem{Straatsma:88}
Straatsma, T.~P.; Berendsen, H. J.~C. {\em J. Chem. Phys.} {\bf 1988}, {\em
  89},  5876.

\bibitem{Ashbaugh:96}
Ashbaugh, H. S.; Wood, R. H. {\bf 1996}, submitted.

\bibitem{Wood:95}
Wood, R. H. {\em J. Chem. Phys.} {\bf 1995}, {\em 103}, 6177.

\end{thebibliography}
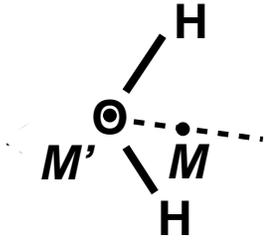
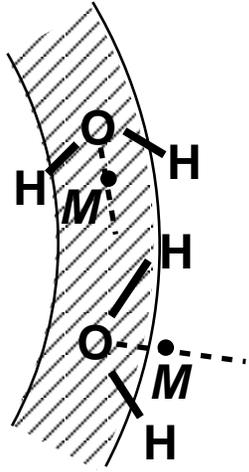
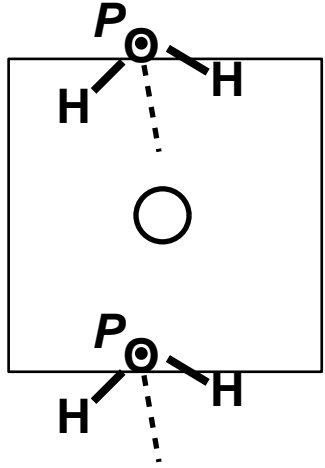
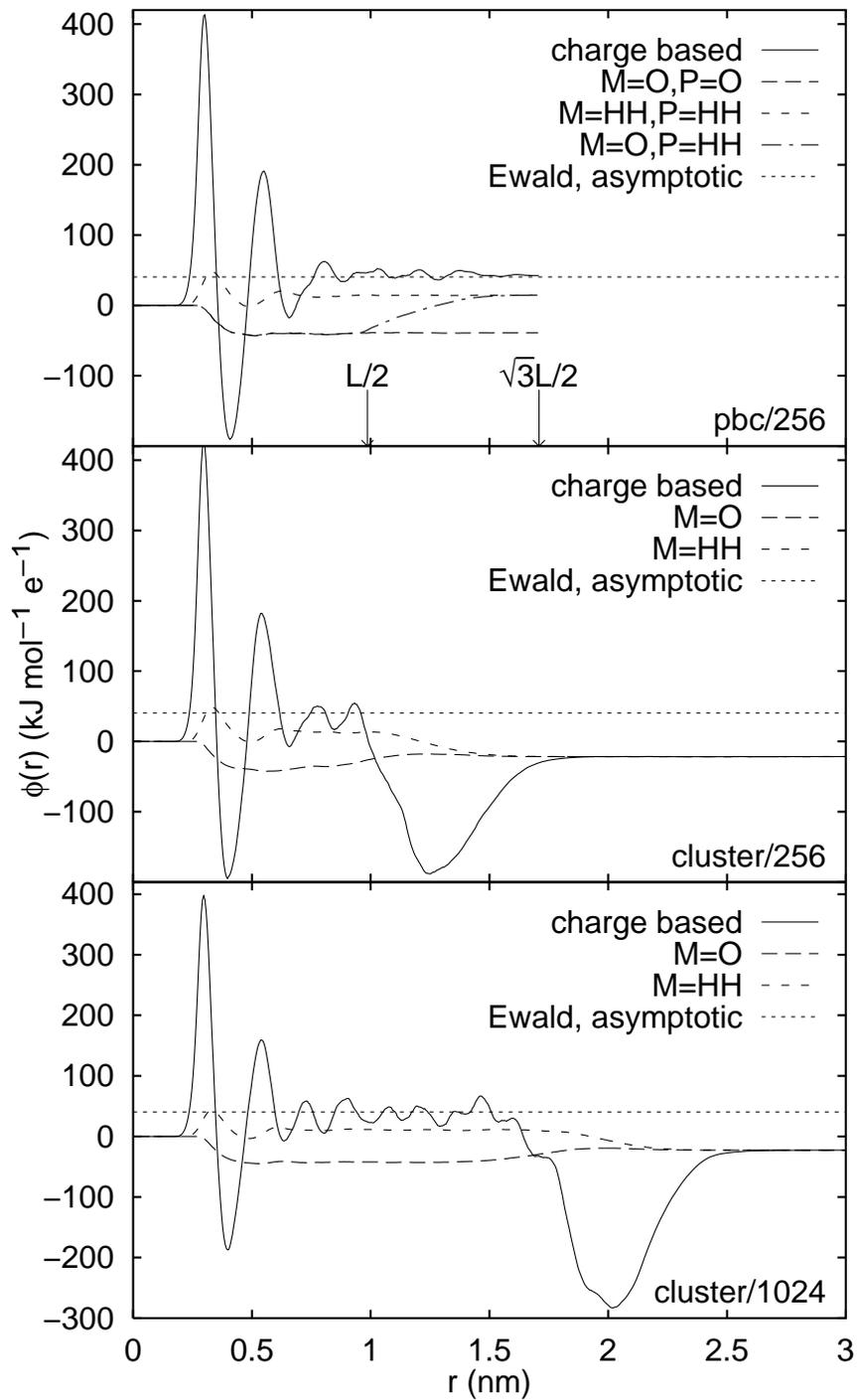
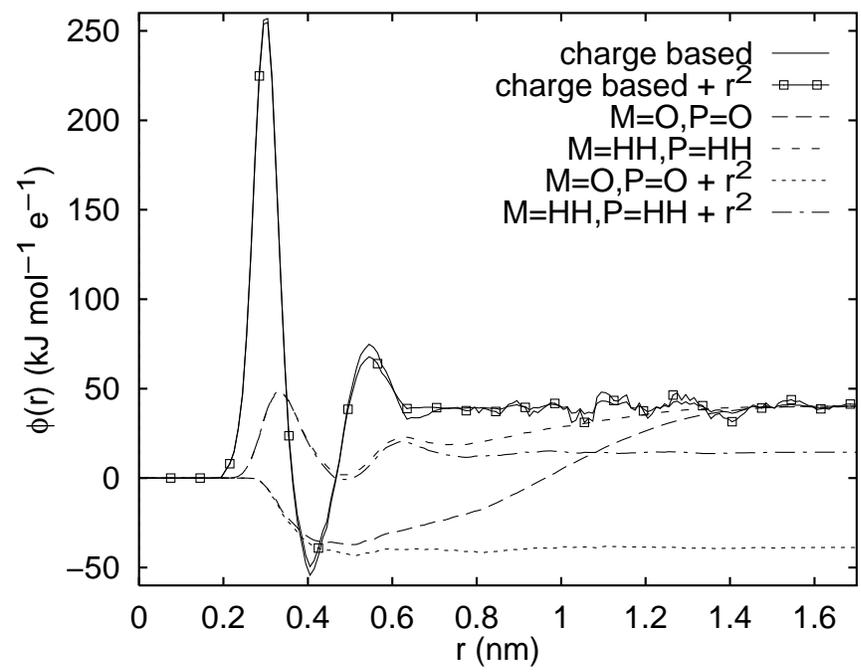
\end{document}